\documentclass[aip,reprint]{revtex4-1}

\usepackage[utf8]{inputenc} 
\usepackage[T1]{fontenc}    
\usepackage{hyperref}       
\usepackage{url}            
\usepackage{booktabs}       
\usepackage{amsfonts}       
\usepackage{nicefrac}       
\usepackage{microtype}      
\usepackage{graphicx}
\usepackage{lipsum}
\usepackage{comment}
\usepackage[caption=false]{subfig}
\usepackage{textcomp}
\usepackage{soul}           
\usepackage{amsmath}
\usepackage{listings}
\usepackage{chngcntr}
\usepackage{enumitem}
\usepackage[table,xcdraw]{xcolor}
\usepackage{booktabs}
\usepackage{multirow}
\usepackage{hhline}

\usepackage{xfrac}
\usepackage{euler}

\definecolor{codegreen}{rgb}{0,0.6,0}
\definecolor{codegray}{rgb}{0.5,0.5,0.5}
\definecolor{codepurple}{rgb}{0.58,0,0.82}
\definecolor{backcolour}{rgb}{0.95,0.95,0.92}

\lstdefinestyle{mystyle}{
    backgroundcolor=\color{backcolour},   
    commentstyle=\color{codegreen},
    keywordstyle=\color{magenta},
    numberstyle=\tiny\color{codegray},
    stringstyle=\color{codepurple},
    basicstyle=\ttfamily\footnotesize,
    breakatwhitespace=false,         
    breaklines=true,                 
    captionpos=b,                    
    keepspaces=true,                 
    numbers=left,                    
    numbersep=5pt,                  
    showspaces=false,                
    showstringspaces=false,
    showtabs=false,                  
    tabsize=2
}
\newcommand{\paragraphtitle}[1]{\textsf{\textbf{\small {#1}}}}

\draft 

\begin{document}

\title[]{Structure and polymerization of liquid sulfur across the $\lambda$-transition } 



\author{Manyi Yang}
\affiliation{Atomistic Simulations, Italian Institute of Technology, 16156 Genova, Italy}

\author{Enrico Trizio}
\affiliation{Atomistic Simulations, Italian Institute of Technology, 16156 Genova, Italy}
\affiliation{Department of Materials Science, Università di Milano-Bicocca, 20126 Milano Italy}

\author{Michele Parrinello*}
\email[]{michele.parrinello@iit.it}

\affiliation{Atomistic Simulations, Italian Institute of Technology, 16156 Genova, Italy}


\date{\today}

\begin{abstract}
    The anomalous $\lambda$-transition of liquid sulfur, which is supposed to be related to the transformation of eight-membered sulfur rings into long polymeric chains, has attracted considerable attention. 
    However, a detailed description of the underlying dynamical polymerization process is still missing. 
    Here, we study the structures and the mechanism of the polymerization processes of liquid sulfur across the $\lambda$-transition as well as its reverse process of formation of the rings. 
    We do so by performing \textit{ab-initio-quality} molecular dynamics simulations thanks to a combination of machine learning potentials and state-of-the-art enhanced sampling techniques. 
    With our approach, we obtain structural results that are in good agreement with the experiments and we report precious dynamical insights into the mechanisms involved in the process. 
\end{abstract}


\maketitle 

\section*{Introduction}
    Elemental sulfur exhibits a very complex phase diagram. This richness derives from the sulfur's propensity to be twofold coordinated. This results in the possibility of either forming ring-like structures ($S_\pi$) or polymeric chains ($S_\infty$) of competing energy. The ring-like arrangements dominate the stable solid  structures~\cite{steudel2003elemental, meyer1976elemental,crapanzano2005alternating} (orthorhombic $\alpha$-S, monoclinic $\beta$-S and $\gamma$-S) with polymeric arrangements reported only for pressures higher than 2.0 GPa~\cite{geller1966pressure,lind1969structure, Crichton2001}. Among the cyclic polymorphs, small rings are preferred, and the 8-membered crown-shaped rings ($S_8$) are by far the most stable configurations. However, evidence of a minority fraction of rings with sizes ranging from 6 to 32 atoms has also been reported~\cite{Henry2020,Bellissent1990}. 
    
    A mixture of these two structural models has also been proposed to describe the liquid phase, and several liquid-liquid phase transitions have been reported in a wide temperature and pressure range~\cite{steudel2003elemental,Steudel2012,meyer1976elemental,Henry2020,liu2014liquid}
    For instance, a transition between a low-density liquid, richer in rings, and a high-density liquid, richer in polymers, has been observed with a transition line that terminates in a critical point at around 1000 K and 2.0 GPa~\cite{Henry2020}.  
    
    Here, we focus on a small part of the phase diagram, namely the in proximity of the so-called \emph{lambda transition}~\cite{Sauer1967,Steudel2012,meyer1976elemental} that occurs at $T_\lambda=$ 432 K and ambient pressure and results in an anomalous behavior of physical properties like heat capacity~\cite{meyer1976elemental}, viscosity~\cite{Sauer1967}, and density~\cite{zheng1992density}. 
    The behavior observed around $T_\lambda$ has been widely studied and associated with the onset of a living polymerization of the $S_8$ rings~\cite{tobolsky1959polymerization, Kozhevnikov2004}. 
    It is believed that before the $\lambda$-transition, ring-like structures dominate, but as one increases the temperature, the polymeric fraction of the liquid slowly grows and suddenly increases around $T_\lambda$.  
   
    Over the years, extensive experimental studies have been conducted to characterize the species involved in the transition~\cite{winter1990,stolz_1994,biermann1998structural,steudel2003elemental,kalampounias2003raman}. However, they still provide limited insight into the underlying dynamical process and a detailed picture of this phenomenon is still missing. Previous \textit{ab initio}-based theoretical studies~\cite{FloresRuiz2022,jones2003dft,Munejiri2000,tse1999dft,wong2002} have partially filled this gap, but the computational costs of first principles methods, which are needed to faithfully describe the complex changes in the forming and breaking of chemical bonds, have limited the scope of these simulations. 
    Indeed, such calculations are still computationally too expensive to perform extensive sampling, even on small systems.
    Thus, all these studies have been limited to relatively short timescales (some 100 ps) and/or simulation cells with only a few hundred atoms. Such limitations are quite severe when studying polymeric systems, which by definition involve a large number of atoms and often exhibit slow dynamics~\cite{hiemenz2007polymer,Kozhevnikov2004}. 
    
    \begin{figure*}[t!]
        \centering
        \includegraphics[width=0.8\linewidth]{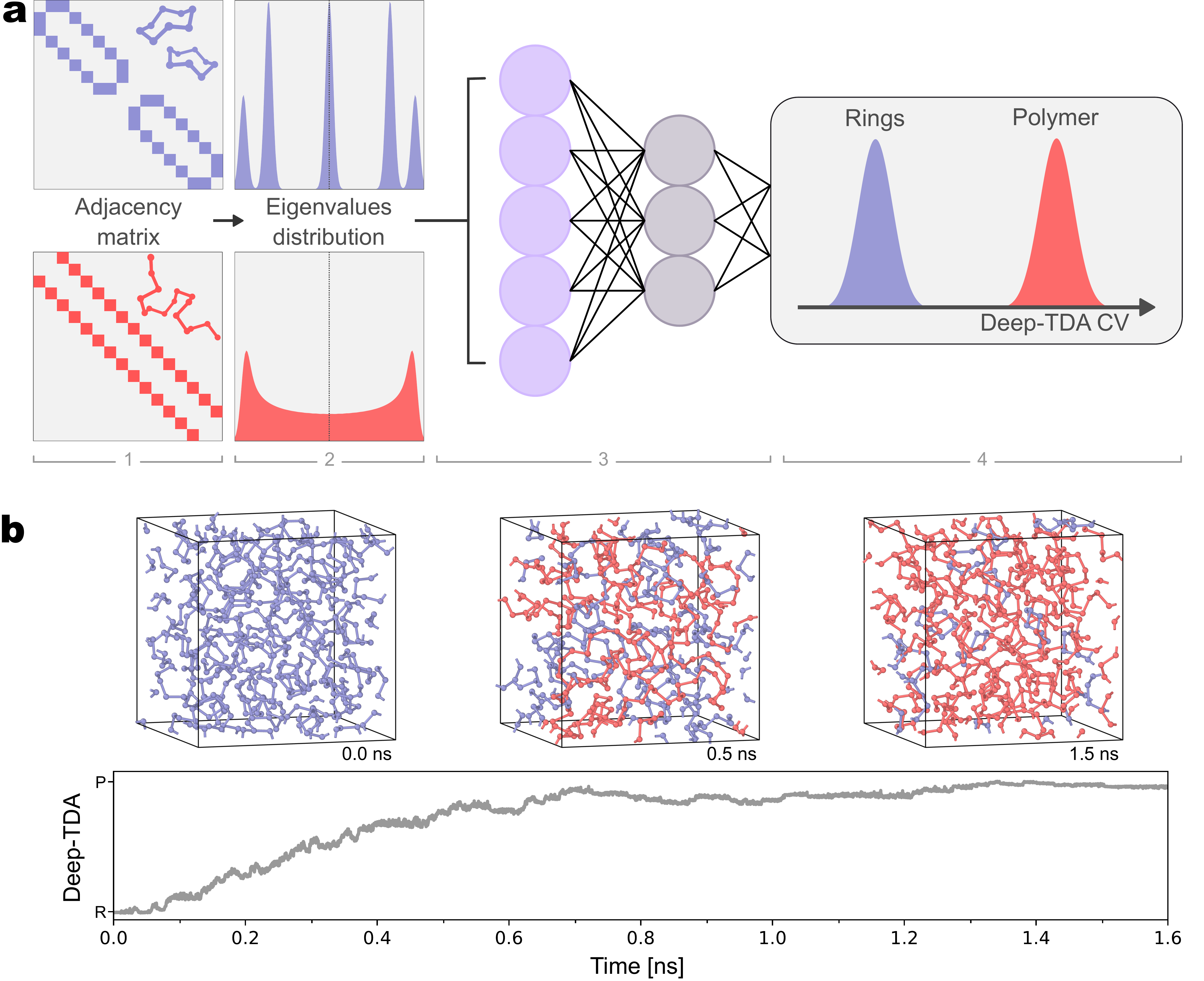}
        \caption{\textbf{a} Schematic representation of the construction of the topological collective variable (CV) for polymerization in liquid sulfur. For each configuration, we build the corresponding adjacency matrix and compute its eigenvalues distribution with a continuous histogram. The values of such histograms are fed as inputs of a neural network (NN) that combines them and returns the CV as output, according to the Deep-TDA CV scheme~\cite{trizio2021enhanced}. For the training of the NN, we build a dataset of configurations from pure rings (blue) and pure polymer (red) phases. The NN is optimized such that the projection of the training data in the CV-space matches a pre-assigned target distribution in which the states are well-discriminated. 
        \textbf{b} Value of the Deep-TDA CV as a function of time in a biased simulation, in which the biasing is applied along the Deep-TDA CV. Starting from a pure $S_8$ phase at 0.0 ns, the system polymerizes to progressively higher polymer fractions. Instantaneous snapshots at times of 0.0 ns, 0.5 ns and 1.5 ns are given.
        In both panels, ring-related features are colored in blue and polymer-related ones in red, respectively.}
        \label{fig:cv_workflow}
    \end{figure*}

    In recent years, the combination of machine-learning interatomic potentials (MLPs) of \emph{ab-initio-quality} and enhanced sampling (ES) in an active learning framework has proven effective in overcoming similar difficulties in processes as complex as the liquid-liquid transition in phosphorus~\cite{yang2021liquid}, the nucleation and phase diagram of gallium~\cite{niu2020gallium}, the decomposition of urea~\cite{yang2022urea} or the dynamical nature of heterogenous catalytic processes~\cite{bonati2023nitrogen, yang2022ammonia, tripathi2023poisoning}.
    In our case, such an approach has allowed the limitation of standard molecular dynamics (MD) simulations to be overcome and to simulate systems of thousands of atoms for timescales of the order of nanoseconds.
    
    Indeed, MLPs allow performing \emph{ab-initio-quality} MD simulations at a fraction of the cost of first principles methods. 
    In this approach, following the strategy pioneered by Behler and Parrinello~\cite{Behler2007}, the interatomic interactions are modeled through a neural network (NN), which is trained to faithfully predict the energies and forces obtained with DFT calculations on a set of reference configurations. 
    To be accurate for the study of reactive processes, it is then of the utmost importance that the training dataset does not contain only low-energy configurations coming from the sampling of metastable states but also includes transition state configurations.
    Unfortunately, in the case of complex systems such as liquid sulfur, due to the presence of large free energy barriers, most reactive events take place on a temporal scale that far exceeds that accessible in standard MD simulations and cannot be sampled.   
    
    Fortunately, ES methods are aimed at overcoming this limitation and allowing rare events to be sampled in an affordable computational time. 
    Many such methods are based on the addition of an external \emph{bias} potential, which is taken to be a function of a small number of collective variables (CVs). 
    The CVs are, in turn, functions of the atomic coordinates  $s = s(\mathbf{R})$ and should be wisely chosen to encode the relevant slow modes of the process for a successful simulation. 
    Recently, the advent of machine learning (ML) methods has greatly facilitated the CV  determination. In the present case, the complex structural changes that occur close to the $\lambda$-transition are difficult to express in simple geometrical terms, and we use instead a more abstract CV that results from a combination of graph theory and ML.
    
    In the first part of the paper, we construct the interaction potential and show that the predictions of our model potential are in agreement with experimental diffraction data and diffusion properties.
    In the second part, we study the polymerization and depolymerization processes that take place close to the $\lambda$-transition. 
    In nanosecond-long reactive simulations, with the help of an analysis of charge distribution, we describe reaction mechanisms that shed light on the puzzling mechanism of formation and breaking of the sulfur polymers.

\begin{figure*}[t]
    \centering
    \includegraphics[width=0.9\linewidth]{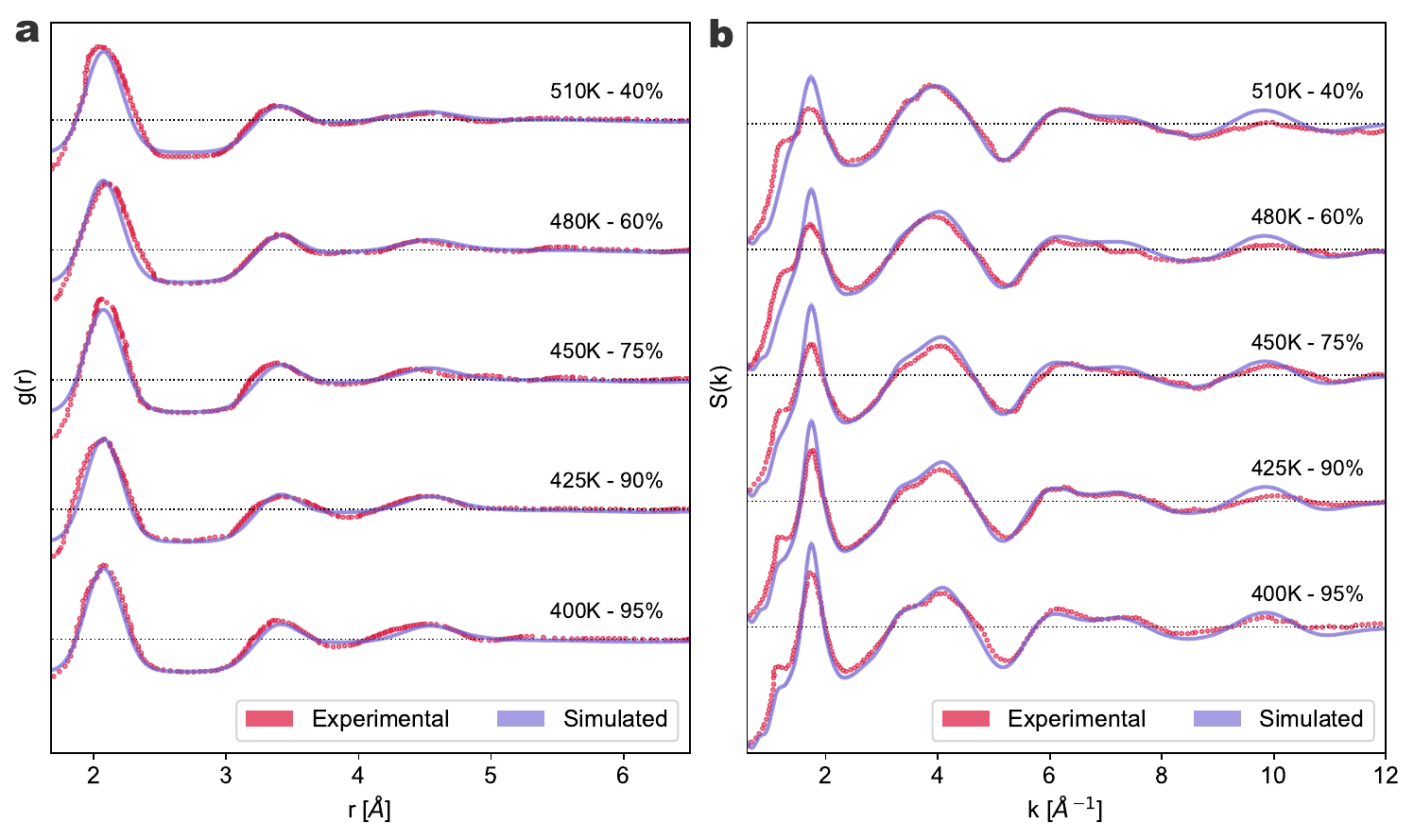}
    \caption{Radial distribution function $g(r)$ (\textbf{a}) and structure factor $S(k)$ (\textbf{b}) for liquid sulfur at temperatures around the $T_\lambda$. 
    Simulated results at different ring concentrations (solid purple lines) are compared with experimental data~\cite{Vahvaselkä_1988} at different temperatures (red void dots). Each experimental temperature and the matching percentage of rings are reported close to the corresponding curve.
    The $\pm 5\%$ interval on the ring concentration for each simulated curve is given as a shaded purple area. 
    The black dotted lines mark the $g(r) = 1$ and $S(k) = 1$ value for each couple of curves as they are offset in the vertical direction for visualization purposes by four and one units, respectively.
    }
    \label{fig:gr_sk}
\end{figure*}
   
\section*{Results}   
    \subsection*{Topological collective variables for enhanced sampling}

        To build the CV for ES simulations, in this work, we use the Deep Targeted Discriminant Analysis~\cite{trizio2021enhanced} (Deep-TDA) method, and the whole protocol of our CV design process is schematically depicted in Fig.~\ref{fig:cv_workflow}\textbf{a}.
        In Deep-TDA, the CV is expressed as the output of a NN that is optimized according to a classification criterion (panels \textbf{3} and \textbf{4}) and takes as input a set of (physical) descriptors, which, in our case, are built from the adjacency matrix of the system (panels \textbf{1} and \textbf{2}). 
        
        As mentioned in the introduction, the use of ML greatly simplified the CV design procedure. However, the effectiveness of such approaches is still affected by the choice of input descriptors, as they should be informative on the slow modes of the process.  
        In this sense, the complex processes that take place at the  $\lambda$-transition presented us with the new challenge of finding proper descriptors that could describe the ring opening and formation while retaining permutational invariance.  
        Since the system undergoes enormous changes in connectivity (see upper panel in Fig.~\ref{fig:cv_workflow}\textbf{b}) we resort to graph theory and, in particular, to the adjacency matrix eigenvalues (panels \textbf{1} and \textbf{2} in Fig.~\ref{fig:cv_workflow}\textbf{a}), which directly relates to the topology of the system. 
        For example, the values $\pm\sqrt{2}$ and $0$ reflect $S_8$ ring arrangements, and the multiplicity of such eigenvalues is proportional to the number of such rings in the system.
        
        In practice, to train our Deep-TDA CV, we used a two-state model, using unbiased data collected in the pure ring and pure polymeric phases (see Fig.~\ref{fig:cv_workflow} whole).
        Our choice is motivated by the experimental evidence that suggests that the relevant properties in the $\lambda$-transition region depend on the relative fraction of these two phases.
        We also note that even if the pure polymeric phase is not reported in the experiments, it still can be used to simplify the training of the model by making the relevant polymer-related features more evident and deepen our understanding of the system, providing a measure of the phase composition of the system during our simulations as shown in the bottom of Fig.~\ref{fig:cv_workflow}\textbf{b}.
        
    \subsection*{Radial distribution function and structure factor}
        The radial distribution function $g(r)$ and the structure factor $S(k)$ depend on the structural ordering of the atoms and their features on the relative concentration of the phases in the sample.
        For this reason, these quantities have been monitored at different temperatures around the $\lambda$-transition~\cite{Vahvaselkä_1988}.
        However, the experimental determination of the $S_8$ fraction is still subject to much uncertainty~\cite{Steudel2012, klement1970, Kozhevnikov2004, crapanzano2006thesis}.  
        
        This is the typical scenario in which theoretical modeling can provide helpful insights.
        Simulations can access the \emph{pure} phases (i.e., only rings $S_8$ and only polymers $S_\infty$), which are never found in the experiments. 
        Despite sounding somehow unphysical, this information can be used to interpret the experimental results.
        One can then obtain the $g(r)$ and $S(k)$ for all the intermediate compositions through a linear combination of the contribution from the pure $S_8$ and $S_\infty$ phases. 
        For example, the radial distribution function $g_\alpha(r)$ at a given concentration of rings $\alpha$ is
        \begin{equation}
            g_\alpha (r) = \alpha g_{S_8}(r) + (1 - \alpha) g_{S_\infty}(r)
        \end{equation}
        and similarly the structure factor $S_\alpha(k)$
        \begin{equation}
            S_\alpha(k) = \alpha S_{S_8}(k) + (1 - \alpha) S_{S_\infty}(k)
        \end{equation}        

        The values of $g_{S_8}$, $g_{S_\infty}$, $S_{S_8}$, $S_{S_\infty}$ were obtained by analyzing 200 configurations collected over one ns of equilibrated dynamics of 3456 atoms (box size 46.9 \AA) in the pure $S_8$ and $S_\infty$ phases.
        We have tested the accuracy of this approach by comparing it with the results from simulations performed at preassigned $S_8$ concentrations, as we report in the SI.  
            
        In Fig. \ref{fig:gr_sk}, we compare the experimental data~\cite{Vahvaselkä_1988} from X-Ray diffraction (XRD) at ambient pressure and different temperatures with our estimates of the $g(r)$ and $S_k$ (panel \textbf{a} and \textbf{b}, respectively) for different $S_8$ concentrations (95\% to 40\% $S_8$ rings).
        The values of $\alpha$ were chosen to best fit the experimental data within the range of concentrations reported in the experiments at the considered temperatures.
        In the case of the $g(r)$, for a meaningful and direct comparison with the experimental data, we also applied a Gaussian convolution to the simulated results. The non-processed data are available in the SI.

        From Fig.~\ref{fig:gr_sk}, it can be seen that the short-range order of the liquid is well-fitted, giving credibility to our estimation of the $S_8$ fraction.
        Remarkably, our results reproduce well the evolution with the temperature of the $g(r)$ third peak around 4.5 $\AA$. This elusive signal is associated with third neighboring distances in the $S_8$, thus tends to disappear as the temperature and the polymer fraction increase.           
        The long-range structure is also well reproduced, except for the pre-peak in $S(k)$ at $k\sim1${\AA}$^{-1}$, which is less apparent in our simulations. 
        This is a reflection of the fact that our system is still too small to resolve this peak well. However, in the previous \emph{ab initio} calculations that out of necessity were performed on smaller cells, both the peak at $k\sim2${\AA}$^{-1}$ and the corresponding pre-peak were not reproduced almost at all.

    \subsection*{Atomic mobility: displacement analysis}\label{subsec:displacements}
        One of the main features of the $\lambda$-transition is its sudden increase in viscosity above $T_\lambda$.
        At the atomic level, this should correspond to a decreased mobility of the atoms.

        Especially in the polymeric phase, the atomic motions become sluggish and a direct calculation of the viscosity is impossible. 
        However, in order to have an insight into the dynamics, we compute and compare the atomic displacements after 10, 25, 50, and 100 ps for ring concentrations that resemble conditions below $T_\lambda$ (100\% rings concentration), slightly above (75\%) and well above (55\%).
        
\begin{figure}[h!]
            \centering            
            \includegraphics[width=1\linewidth]{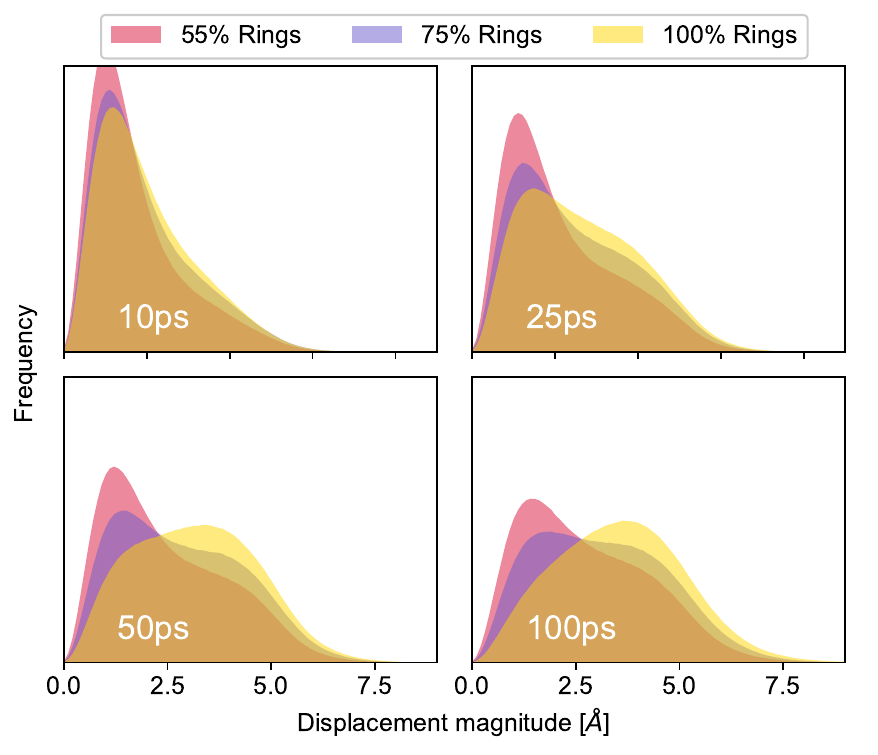}
            \caption{Histogram of the atomic displacements in liquid sulfur at different ring concentrations, given in the legend, and different lag times, indicated by white labels.}
            \label{fig:displacements}
        \end{figure}

        The results from this analysis are reported in Fig.~\ref{fig:displacements} and clearly show that, on average, atoms in the molecular $S_8$ phase have the highest mobility. 
        On the other hand, as the polymeric content increases, a peak in the distribution starts to appear below the $\sim$2{\AA} threshold of the first coordination shell. 
        This comes from the polymer atoms, which mostly oscillate around their positions rather than showing any net drift, thus inducing the rise in the viscosity.
    
    \subsection*{Reaction mechanisms and charge analysis}      
        Having assessed the reliability of our potential's prediction when compared to the experiments, we move to the study of the chemical mechanisms involved in the $\lambda$-transition with the crucial help of enhanced sampling simulations and our topological CV. In the following, we propose mechanisms for the polymerization/depolymerization process.
        We note that we report only those mechanisms that we found to be dominant in our simulations, i.e., they were observed in the majority of several independent simulations, and for all of them, we double-checked the agreement of our potential with DFT calculations to avoid artifacts.

        For each mechanism, we provide a prototypical example and an analysis of the instantaneous charges involved in the process. This analysis is made with the help of Bader charge distribution~\cite{henkelman2006fast} as obtained from the DFT charge density.

        A schematic chemical diagram for the proposed mechanism is also available in the SI.
        
        \begin{figure*}[!ht]
            \centering
            \includegraphics[width=\linewidth]{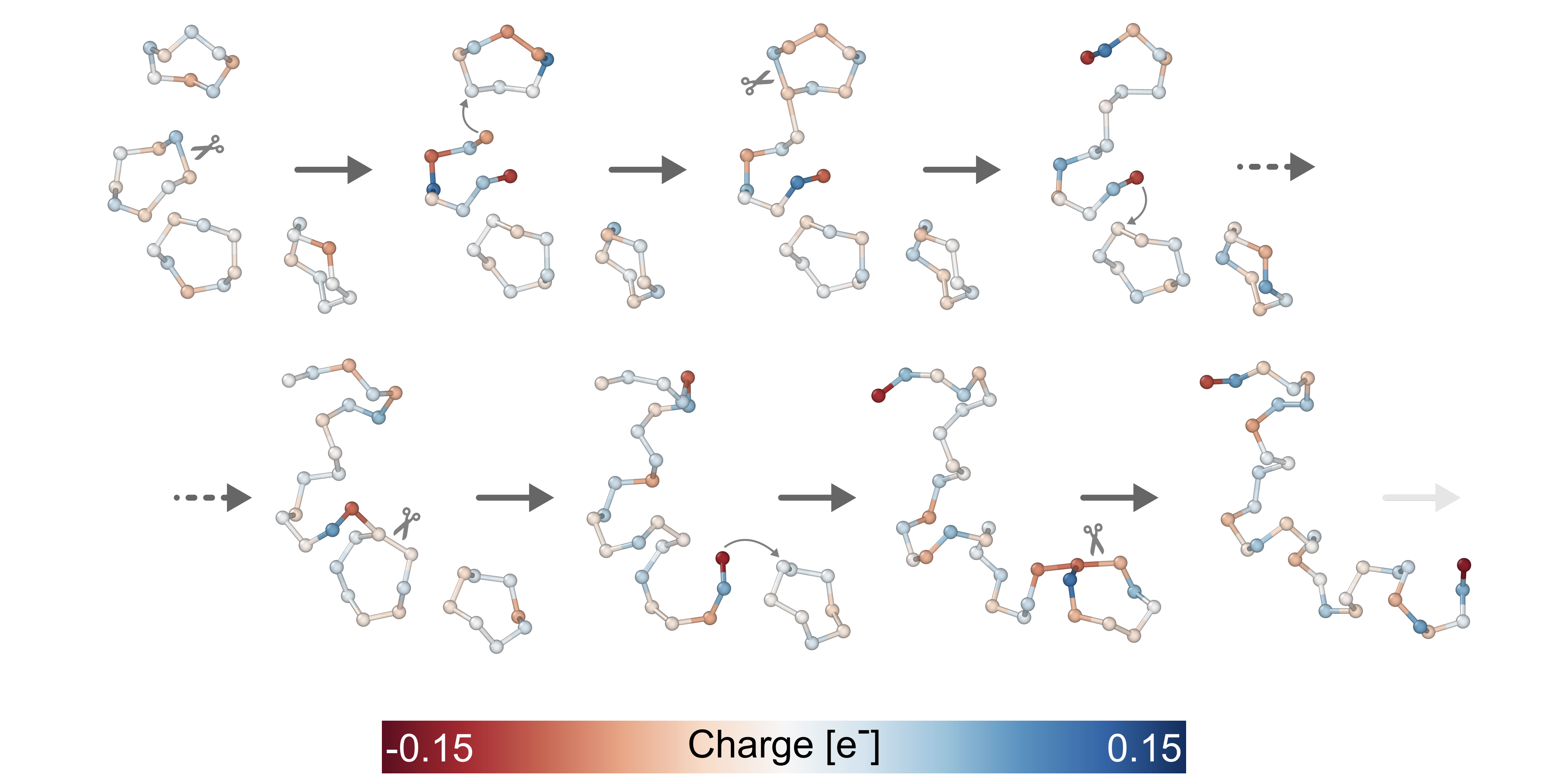}
            \caption{Polymerization mechanism in liquid sulfur starting from four $S_8$ rings. The atoms are colored according to the instantaneous charge obtained by computing the Bader's charge from the DFT electronic density. For visualization purposes, only the relevant atoms are represented from the 512 in the simulation cell.}
            \label{fig:polymerization}
        \end{figure*}        
            
        \paragraphtitle{Polymerization mechanism}
            The polymerization reaction in liquid sulfur, which we schematically depict in Fig.~\ref{fig:polymerization}, resembles that of the standard \emph{ring-opening} polymerization. 
            The first step requires the formation of an active center from which the polymerization can propagate.
            The active center forms when one of the crown-shaped $S_8$ monomeric units undergoes such large thermal fluctuations that it manages to open. 
            The ring deformation induces a charge polarization, as shown in Fig.~\ref{fig:polymerization}. 
            Negative charges concentrate on the under-coordinated terminal atoms, thus making them active. At this point, they can either react together to close again the ring, or they can look for new neighbors on a different ring nearby.
            As the active terminal interacts with the guest ring, its charges are forced to reorganize. This induces a deformation in the guest ring that may eventually open it, leading to the formation of the first oligomer.
            Right after the opening of the second ring, the charges along the new short chain quickly reorganize, and negative charges concentrate in the chain tails. 
            This makes the terminals active again and capable of further propagating the polymerization.

            Even if the ideal crown-shaped $S_8$ rings dominate the liquid phase, our calculations confirmed the experimental evidence~\cite{steudel2003elemental} that other cyclic monomers ($S_n$, n $\neq$ 8) and sub-stable isomers of $S_8$ rings (see SI) also contribute to this polymerization process.
            
            Overall, from our results, it clearly appears that the under-coordinated nature of the chain tails mainly drives the polymerization. This makes the terminal atoms highly reactive, as indicated by the negative charge localization, and eager to find new partners. 

            Another key factor is related to the increased stabilization of the charge unbalance in longer polymeric strands. In that case, the average delocalization of the charge is more efficient with respect to the case of short oligomers.  
            In this regard, we also found, in agreement with previous theoretical studies~\cite{jones2003dft}, that shorter chains tend to be rather unstable and often revert to rings if left to relax in short unbiased dynamics, at variance with the longer chains, which remain stable.
        
\begin{figure*}[!ht]
              \centering
              \includegraphics[width=0.9\linewidth]{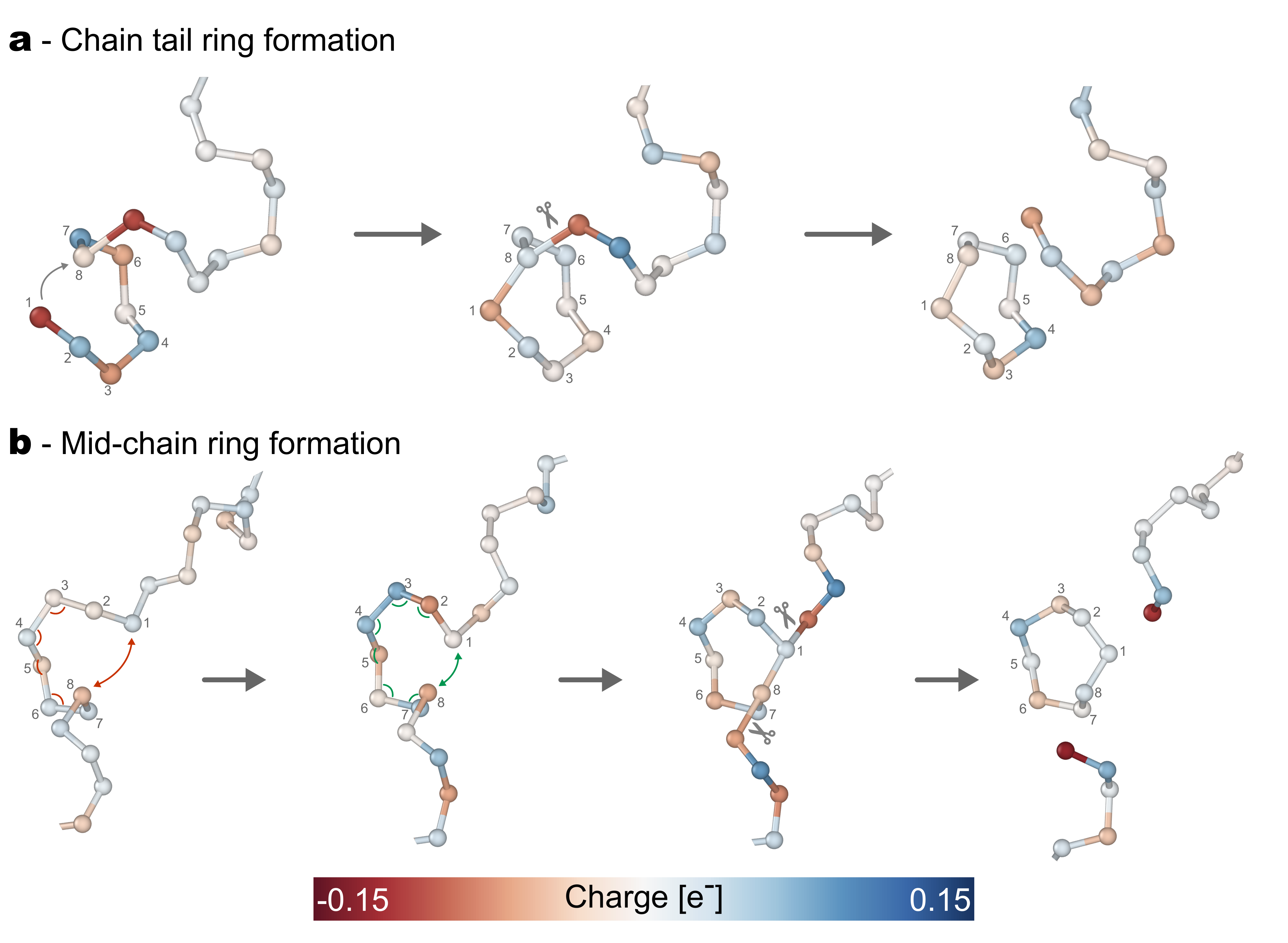}
              \caption{Mechanisms for the formation of rings in liquid sulfur starting from $S_\infty$ polymers. \textbf{a} Formation of a ring from the tail of the polymeric chain. \textbf{b} Formation of a ring in the middle of the chain.
              The atoms are colored according to the instantaneous charge obtained by computing the Bader's charge from the DFT electronic density.
              For visualization purposes, only the relevant atoms are represented from the 512 in the simulation cell.}
              \label{fig:ring_formation}
            \end{figure*}

        \paragraphtitle{Ring formation mechanisms}
            The formation of the rings starting from the polymer can occur either at the end of the chain (see Fig.~\ref{fig:ring_formation}\textbf{a}), as one would intuitively suppose, but somehow surprisingly, also in the middle, as we schematically depict in Fig.~\ref{fig:ring_formation}\textbf{b}.
            
            In the first case (see Fig.~\ref{fig:ring_formation}\textbf{a}), the chain tail is characterized by two elements: a charge unbalance and higher mobility with respect to the rest of the chain.
            The first element is specifically due to the sulfur atoms being under-coordinated, thus leading to their negative polarization. 
            This is the \emph{chemical} driving force of the reaction, as it makes such atoms eager for new partners and ready to react.
            However, this is not enough for the reaction to take place, as the reactive tail has to fold onto the chain to form the loop that will eventually lead to the ring.
            This is made possible thanks to the higher mobility of the terminal atoms, which is typical of polymeric chain ends.
            Of course, this loop is most stable if the terminal atom folds such that it interacts with its 7$^{th}$ neighbor, thus ensuring the $S_8$ arrangement, but this same mechanism can also lead with less probability to some of the different-sized rings as we report in the SI. 
            
            In the second mechanism, the scenario is significantly different as the atoms involved in the formations of the loop belong to the middle of the polymeric chain (see Fig.~\ref{fig:ring_formation}\textbf{b}).
            Such atoms are indeed fully coordinated, meaning that they are much less reactive than in the first case and that any polarization fluctuation is short-lived. 
            We found that to compensate for this weaker reactivity, it is crucial that the arrangement of the atoms resembles as much as possible that of one of the stable $S_\pi$ rings. Of course, the choice shall preferably be the ideal crown-shaped $S_8$. Thus, we report this case in Fig.~\ref{fig:ring_formation}\textbf{b}.
            In the first frame, the eight-membered loop that starts to appear in the middle of the chain still has a wrong combination of angles and distances. 
            On the other hand, in the configuration reported in the second frame, the sequence of angles becomes favorable, and the distance between the 1$^{st}$ and 8$^{th}$ S atoms reduces enough they can interact. 
            As a consequence of this interaction, the adjacent atoms show a weak negative polarization (see third frame), which becomes stronger as the ring finally separates from the original chain. 
            
\section*{Discussion}
    In this work, we studied the structures and the polymerization and depolymerization mechanisms across $\lambda$-transition in liquid sulfur using a combination of \textit{ab-initio-quality} machine learning potentials and state-of-the-art enhanced sampling techniques.
    For an effective application of the latter approach, we designed a topological collective variable by combining machine learning and graph theory. 
    This has been crucial for the construction of a proper dataset on which our machine-learning potentials were optimized and for the study of the polymerization and depolymerization mechanisms of liquid sulfur. 
    Our methodology proved powerful and greatly improved over previous DFT-based theoretical results. The calculated static structural quantities, including $g(r)$ and $S(k)$, are in good agreement with the experimental results. Similarly, our calculations reproduce well the trend in the atomic mobility in the different phases that can be associated with the anomalous rise in the viscosity during the sulfur polymerization across the $\lambda$-transition.
    In addition, we proposed dynamic reaction mechanisms for the processes involved in the $\lambda$-transition, which also shed light on the critical role that charge localization plays in driving the polymerization and depolymerization process. 

    Overall, our calculations allowed us to harvest precious results in close agreement with the experiments and the success of our strategy encourages us to study other complex systems in the future with this approach.

\section*{Methods}
\subsection*{Code and software}
        All \textit{ab initio} molecular dynamics simulations (AIMD) and single-point energies and forces needed for training the neural network (NN) potentials were performed using the CP2K 9.2 code~\cite{cp2k2020}. In addition, we double-checked that the quantities computed using the CP2K code are consistent with those obtained with Quantum Espresso code~\cite{Giannozzi_2017}.
        The Bader's charges were obtained by analyzing the DFT electronic densities with the Bader~\cite{henkelman2006fast} code.

        The DeepMD-kit package~\cite{wang2018deepmd, zeng2023deepmdkit2} was used for the training of the NN potentials for the atomic interactions and as a plugin in the LAMMPS~\cite{lammps} MD engine for performing the simulations.
        
        For the application of enhanced sampling methods, we relied on the PLUMED~\cite{tribello2014plumed} plugin patched with LAMMPS.
    
        The training of the machine learning collective variable (Deep-TDA CV) has been done using the \texttt{mlcolvar}~\cite{bonati2023mlcolvar} library based on PyTorch~\cite{paszke2019pytorch}. In particular, we refer to the \texttt{4387073} commit\footnote{https://github.com/luigibonati/mlcolvar/commit/4387073} of such a library, which includes additional preprocessing tools for adjacency-matrix-related calculations.
        The CVs have been deployed to PLUMED using the interface provided in its \texttt{pytorch} module~\cite{bonati2023mlcolvar}.
    
        The atomic displacement analysis was performed using the visualization and post-processing code Ovito~\cite{ovito}, which was also used for the rendering of the molecular snapshots reported in the paper.

    \subsection*{AIMD simulations}
        In AIMD simulations, the energies and forces were computed using the Perdew–Burke–Ernzerhof (PBE) exchange-correlation density functional~\cite{PBE}. 
        The Kohn and Sham orbitals were expanded in a m-DZVP Gaussian basis and the plane wave expansion of the electronic density was truncated at an energy cutoff of $300$ Ry. The core electrons were treated using the Goedecker-Teter-Hutter (GTH) pseudopotentials~\cite{goedecker1996separable,hartwigsen1998relativistic} optimized for PBE. 
        To keep the computational cost low, only the $\Gamma$-point was used to sample the supercell Brillouin zone.

        All AIMD simulations were performed in the NPT ensemble with a time step of 2.0 fs. Temperature and pressure were controlled using  Nos\'{e}-Hoover thermostat\cite{evans1985nose} and a  Nos\'{e}-Hoover-like barostat~\cite{melchionna1993hoover} with coupling constants of 0.05 ps and 0.5 ps, respectively. 
        To mitigate the computational costs, only a smaller cubic simulation cell consisting of 128 atoms was used. 
        Nonetheless, the results discussed in the text were calculated on larger cubic simulation cells. 

    \subsection*{Single-point energies and forces calculations}
        The energies and forces needed for the NN potential training were computed using the same exchange-correlation density functional (PBE) and pseudopotential (GTH) as that of the AIMD simulations. However, the energy cutoff was increased to 350 Ry and the D3 dispersion corrections~\cite{D3-grimme2011effect} were included. 
        Single-point calculations have been performed on cells with 128 and 512 atoms.  
        For the smallest cells, we used $k$-points grids of 2$\times$2$\times$2. In contrast, we only used the $\Gamma$-point for the larger ones as we checked that the accuracy on energies and forces 
        with this setup was almost indistinguishable from the one on smaller cells using the grids.

\subsection*{NN potential-based MD simulations}
        All the results reported in the text were obtained using NN-potential-based MD simulations. Specifically, we performed unbiased and biased simulations with a timestep of 1.0 fs in the NVT ensemble, using the global velocity rescaling thermostat~\cite{bussi2007canonical} with a relaxation time of 0.05 ps and periodic boundary conditions (PBC) on cubic simulation cells with box sizes chosen to be consistent with the experimental densities. 
        
        The unbiased approach has been used to study the radial distribution function, the structure factor, and the atomic mobility. For this latter analysis, we simulated a system consisting of 512 atoms (box 24.8\AA) starting from configurations generated from biased simulations, whereas, for the others, we simulated a much larger cell with 3456 atoms (box 46.9\AA). 
         
        The more expensive enhanced sampling approach has been used to simulate the dynamics of 512 atoms to study the polymerization and depolymerization mechanisms of sulfur.

    \subsection*{Enhanced sampling method: OPES}
         We reverted to enhanced sampling to study the polymerization and depolymerization processes of sulfur that are supposed to take place close to $\lambda$-transition. In particular, we used the on-the-fly probability enhanced sampling (OPES) method~\cite{invernizzi2020rethinking,invernizzi2022exploration}. 
         In OPES, the equilibrium probability distribution $P(\boldsymbol{s})$ in the CV space $\boldsymbol{s}$ is estimated on the fly, and a bias potential $V_n(\boldsymbol{s})$ is constructed so as to drive $P(\boldsymbol{s})$ toward a target distribution $P^{tg}(\boldsymbol{s})$.  
        
        Here, we used as target the well-tempered distribution~\cite{bonomi2009welltempered} $P^{tg}(\boldsymbol{s}) \propto [P(\boldsymbol{s})]^{\frac{1}{\gamma}}$, where $\gamma>1$ is the bias factor.
        In this case, the bias potential at $n^{th}$ iteration is written as: 
            \begin{equation}
                   V_n(\boldsymbol{s}) = (1-1/\gamma)\frac{1}{\beta}\log\left ( \frac{P_n(\boldsymbol{s})}{Z_n} + \epsilon \right )
                \label{eq:opes}
            \end{equation}
         where $Z_n$ is a normalization factor, $\beta = 1/{k_BT}$, and $\epsilon = \mathrm{e}^{-\beta \Delta E / (1 - 1/\gamma)}$ is a regularization parameter that controls the maximum deposited bias $\Delta G$. 
         In this paper, we set the frequency for kernel deposition to 250 and $\Delta G$ in the range between 100 and 200 kJ/mol (i.e., \texttt{STRIDE} and \texttt{BARRIER} parameters in PLUMED input files, respectively).   

    \subsection*{Collective variables method: Deep-TDA}
        The OPES biasing potential is applied along a small set of CVs, which are continuous and derivable functions of the atomic coordinates $s = s(\mathbf{R})$.
        These are meant to encode the slow modes of the system and, as an additional requirement, they shall be invariant with respect to the symmetries of the system (rotation, translation, and permutation of identical atoms).
        
        In the Deep Targeted Discriminant Analysis (Deep-TDA) method, the CV is the output of a feed-forward NN (see panel 3 in Fig.~\ref{fig:cv_workflow}\textbf{a}) whose inputs are a set of physical descriptors, such as distances, angels, coordination numbers, collected with short unbiased runs in the metastable basins that are supposed to be visited in the process of interest.       
        The NN is optimized such that the training data, when projected in the CV space, are distributed according to a preassigned target distribution (panel 4 in Fig.~\ref{fig:cv_workflow}\textbf{a}). This target is defined as a series of Gaussians with fixed positions and widths, one for each state, such that data from different basins are localized in different regions of the CV space.

    \subsection*{Topological collective variable training}
        In our Deep-TDA CV model, we used the \verb|mlcolvar| library~\cite{bonati2023mlcolvar} preprocessing tools to compute the NN input descriptors starting from the Cartesian coordinates of all the atoms in the system, which are the inputs of our deployed model in PLUMED.

        To compute the NN input descriptors, we build the adjacency matrix $\mathbf{A}$ in which the $\mathbf{A}_{ij}$ element is $\mathbf{A}_{ij}=1$ if the scalar distance $d_{ij}$ between atom $i$ and $j$ is lower than a cutoff distance $d_{cutoff}$ and $\mathbf{A}_{ij}=0$ otherwise.
        However, the application of such a sharp cutoff would give a matrix $\mathbf{A}$ with discontinuous derivatives that would not be suited for a biasing context.
        Thus, we applied the cutoff using a sharp switching function $S(d_{ij})$ in the form 
        \begin{equation}
            S(d_{ij}) = \frac{1}{1 + \exp(\frac{d_{ij} - d_{cutoff}}{q})} 
        \end{equation}
        where the $q$ value was chosen to obtain the sharpest behavior with numerically stable derivatives, i.e., $q = 0.25$, and $ d_{cutoff}$ was set to 2.6$\AA$ based on the typical sulfur-sulfur bond distances.

        We then computed the full eigenvalues spectrum of $\mathbf{A}$ with Pytorch~\cite{paszke2019pytorch} tools and computed a histogram of such values with 100 bins in the range (-2.2,2.2) using a Gaussian expansion to ensure continuous derivatives.
        Finally, we took the values of the histogram in the 100 bins as input for the Deep-TDA NN. 
        In the training, the Deep-TDA targets were chosen to be $\mu_A=-25$ and $\mu_B=25$ for the centers of the distributions and $\sigma_A=0.2$ and $\sigma_B=0.2$.
        The training set was composed of 18000 configurations for each of the two states for a total of 36000 configurations.
        Including input and output layers, the architecture was [100,64,32,1] nodes/layer, and we used the rectified linear unit (ReLU) as activation function with a learning rate of 0.001 for the optimization. 
   
    \subsection*{NN potential for interatomic interactions training }
        The NN potentials were trained with the DeepMD method~\cite{zhang2018} using the attention-based Deep Potential scheme~\cite{zhang2022dpa}, which, in our case, is able to give a much more accurate reproduction and prediction of DFT energies and atomic forces compared to the standard Deep Potential-Smooth Edition scheme~\cite{zhang2018end} (see SI). 
        The cutoff radius was set to smoothly decay from 0.5 {\AA} to 7.5 {\AA}. The maximum possible number of neighbors in the cutoff was set to 90, and the number of layers in the attention scheme to 3. We used three hidden layers with [30, 60, 120] nodes/layer for the embedding network and four hidden layers with [240, 240, 240, 240] nodes/layer for the fitting network, whereas the size of the embedding matrix was set to 16. 
        The learning rate was set to decay from 1.0 $\times$ 10$^{-3}$ to 5.0 $\times$ 10$^{-8}$ an we used a batch size of 8. 
        The prefactors of the energy and force terms in the loss function were set to change during the training from 0.01 to 5 and from 1000 to 1, respectively, and the final NN model was trained for 3.0 $\times$ 10$^6$ steps.

    \subsection*{Collection of the training set}
        The key step in the construction of a machine learning potential is the collection of the training data set.  This is particularly challenging in the case of liquid sulfur 
        that exhibits a wide range of ring-like and chain-like metastable structures~\cite{steudel2003elemental} .  It is thus necessary to include these configurations in the training set as well as those related to the transition state of the interconversion process.

        To improve and simplify this step, active learning strategies boosted by the use of enhanced sampling methods have been applied to study several complex systems~\cite{bonati2018,niu2020gallium,yang2022urea, yang2021liquid, yang2022ammonia,bonati2023nitrogen}. 
        The advantage of this approach is that it allows crucially relevant reactive configurations to be extracted at an affordable computational cost.

        In our specific case, we applied the On-the-fly probability enhanced sampling ~\cite{invernizzi2020rethinking,invernizzi2022exploration} (OPES) method combined with state-of-the-art machine learning collective variables (CVs) (see topological collective variables for enhanced sampling section).
        The whole procedure of exploring the relevant atomic configurations in the training set is as follows. 
        
        First, we ran a series of unbiased AIMD simulations in the NPT ensemble on systems of 128 atoms for times ranging from 2 to 10 ps. 
        These simulations were performed in the 500 $\sim$ 1200 K temperature range and 0.2 $\sim$ 3.0 GPa pressure range to collect configurations in both the polymeric and ring phases of liquid sulfur. 
        Indeed, approaching high temperature and pressure, $S_8$ rings are destabilized in favor of the $S_\infty$ polymers. 
        
        From these simulations, we collected about 7800 atomic configurations to build the initial training set and start our active learning procedure, which alternate cycles of training and sampling according to the following steps:    
        \begin{itemize}
               \item \textbf{Step 1:} We train four NN potentials using different initial weights and the previous iteration's updated training set. For the first iteration, we shall use the initial training set of AIMD configurations.
        
                \item \textbf{Step 2:} We perform a series of simulations using one of the four NN potentials trained in Step 1 to explore new relevant atomic configurations.
                Not limited to AIMD simulations anymore, we expand our systems from 128 atoms to 512 atoms and run enhanced sampling simulations using OPES combined with our topological CVs. 
                These biased simulations are essential for exploring the active atomic configurations along the polymerization of $S_8$ rings. 
                
                During the simulations, we monitor the reliability of the potential on the sampled configurations based on the maximal standard deviation $\sigma$~\cite{zhang2020dp} of the atomic forces predicted by the four NN potentials:
                    \begin{equation}
                        \sigma=\mathop{max}\limits_{i}\sqrt{\frac{1}{4}\sum_{\alpha=1}^{4}{\Vert{ \boldsymbol{F}_{i}^{\alpha}-\overline{\boldsymbol{F}_{i}}}\Vert^2}}
                    \end{equation}
                where $\boldsymbol{F}_{i}^{\alpha}$ is the atomic force on the atom $i$ predicted by the NN potential $\alpha$, and $\overline{\boldsymbol{F}_{i}}$ is the average force on the atom $i$ over the four NN potentials. 
                   
                To minimize the number of new relevant atomic configurations to be added to the training set while ensuring maximum diversity, we follow the same strategy described in Ref.~\cite{yang2022ammonia}, with the low ($\sigma_l$) and up ($\sigma_u$) bound values set to 0.15 and 0.4, respectively. 
                We refer the readers to Ref. \cite{yang2022ammonia} for further details. 
            
                \item \textbf{Step 3:} We calculate the DFT energies and forces for the configurations selected in Step 2 and include them in the training set for the next iteration.
            \end{itemize}
                 
                Following our previous work, this active learning process is repeated until less than 10$\%$ of the sampled atomic configurations fall into the candidate list of Step 2. 
                At the end of the procedure, our dataset included roughly 1.5 $\times$ 10$^5$ atomic configurations, with almost 90\% of them consisting of 512 atoms.



%
%

%

\section*{Data availability}
    All the inputs and instructions to reproduce the results presented in this manuscript will be made available in the PLUMED-NEST repository upon publication.

\section*{References}
\bibliography{references}

\begin{acknowledgments}
    M.Y. and E.T want to thank Ana Maria Borrego Sanchez, Umberto Raucci, Luigi Bonati for useful discussions and support.
    The authors acknowledge the HPC infrastructure and the Support Team at Fondazione Istituto Italiano di Tecnologia. Computational resources were also provided by the Swiss National Supercomputing Centre (CSCS) under project ID $S1134$ and $S1183$.
\end{acknowledgments}

\section*{Author contributions Statement}
    M.Y., E.T., and M.P. made substantial contributions to the design and implementation of the work and to the writing of the manuscript.  

\section*{Competing Interests Statement}
    The authors declare no competing interests.

%
%
\end{document}


\maketitle
\newpage
\tableofcontents
\newpage

\section{Supplementary Discussion}

    \subsection{Radial distribution function and structure factor}
\begin{figure}
            \centering
            \begin{minipage}{0.49\linewidth}
                \includegraphics[width=\textwidth]{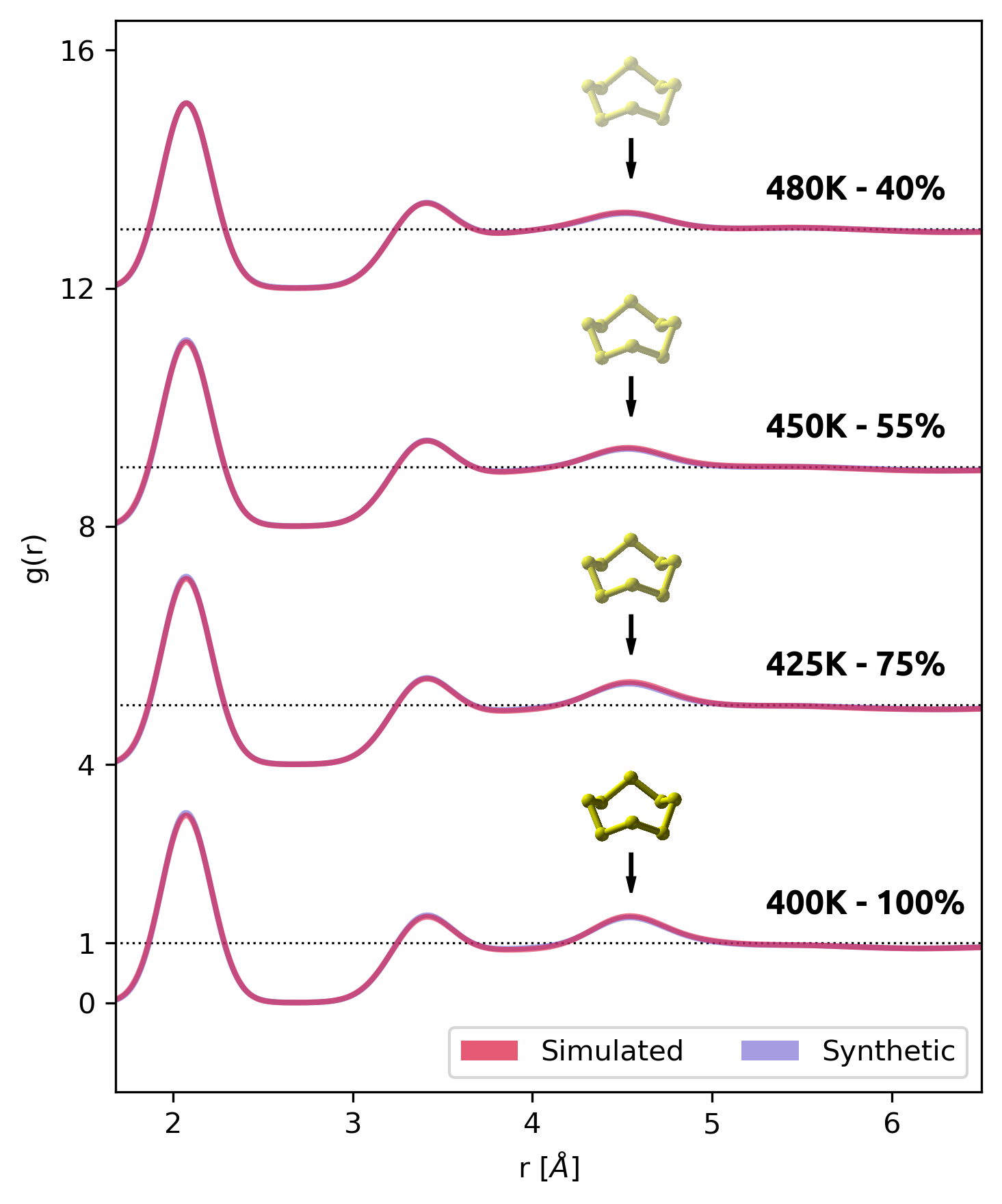}
            \end{minipage}
            \hspace*{\fill}
            \begin{minipage}{0.49\linewidth}
                \includegraphics[width=\textwidth]{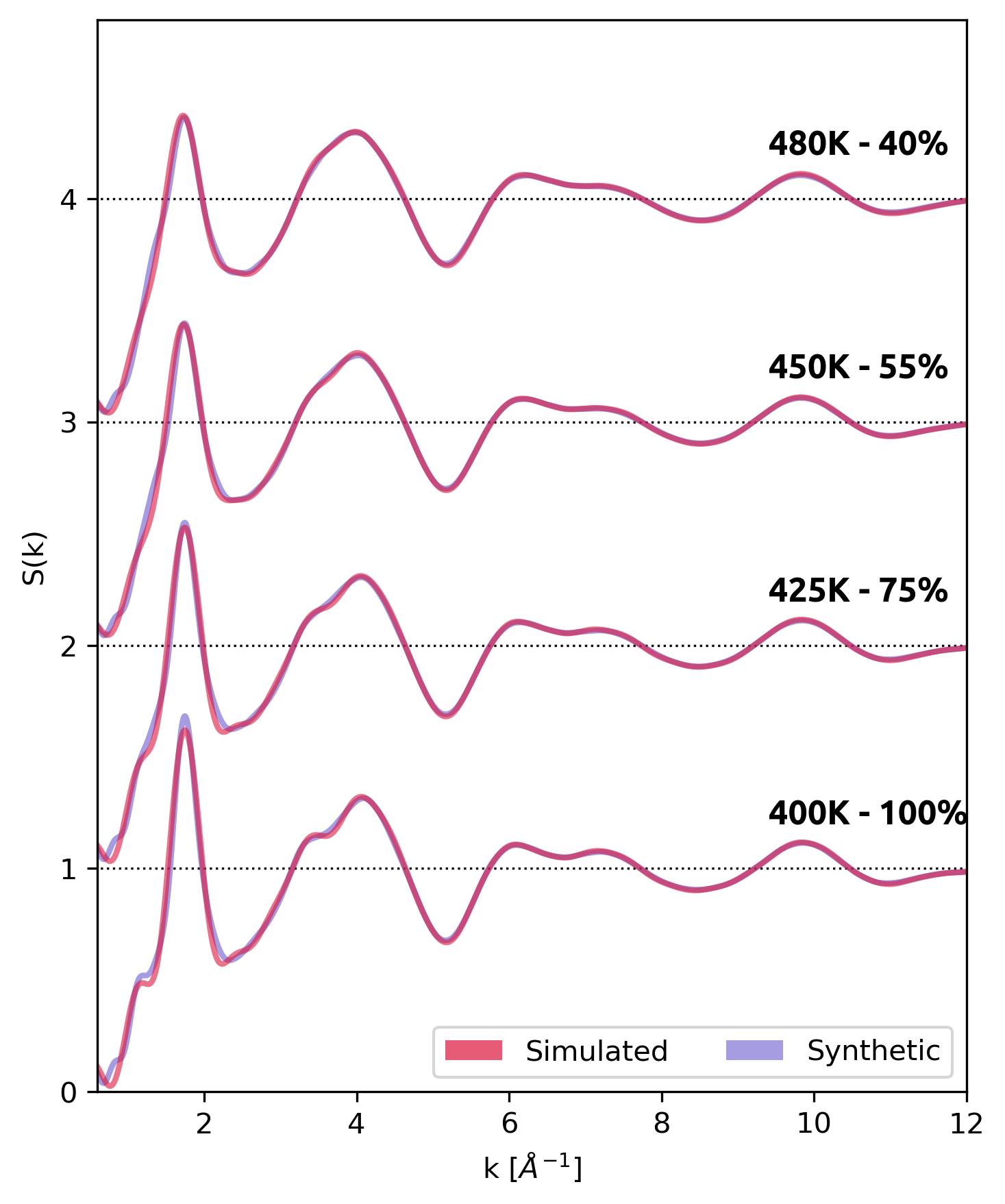}
            \end{minipage}
            \caption{Comparison of the radial distribution function and structure factor obtained using the approach reported in the main text for a 3456 atoms system and from unbiased simulations at the corresponding concentration of rings with a 512 atoms simulation cell.}
            \label{supp_fig:rdf_sk_syn_sim}
        \end{figure}

        In the main text, we report the $g(r)$ and the $s(k)$ for different ring concentrations generated with a \emph{synthetic} approach starting from the same quantities computed for the pure phases in larger cells of 3456 atoms. 
        To validate this approach, we report in Fig.~\ref{supp_fig:rdf_sk_syn_sim} these quantities computed from simulations at the corresponding concentrations on a smaller cell of 512 atoms. 
        It is evident that the two results are almost indistinguishable.
     
\begin{figure}[!h]
            \centering
            \includegraphics[width=0.5\textwidth]{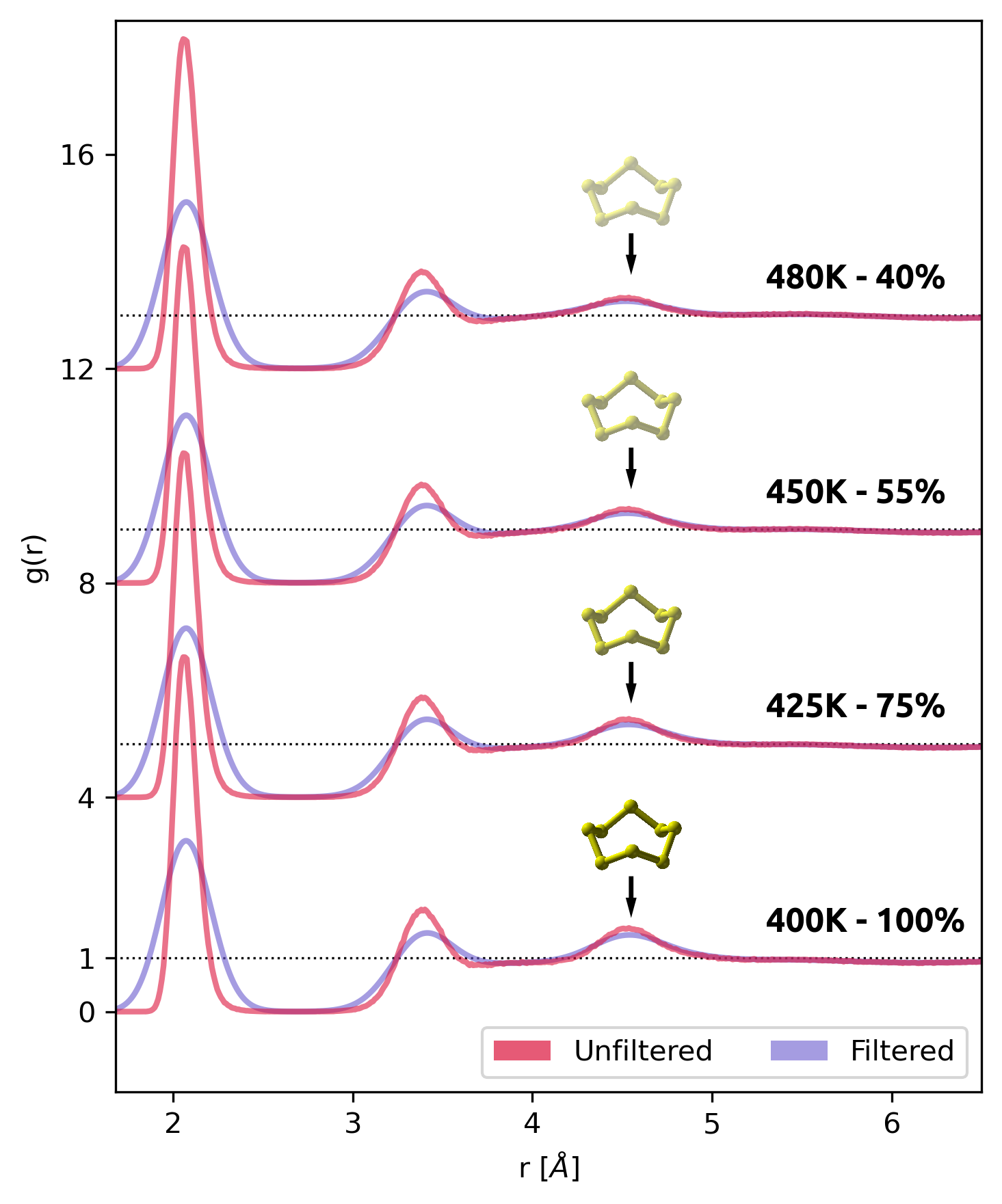}
            \caption{Comparison of the radial distribution function before and after the application of the Gaussian filtering.}
            \label{supp_fig:rdf_filter}
        \end{figure}

\clearpage      
       In Fig.~\ref{supp_fig:rdf_filter}, we also report the $g(r)$ curves reported in the main text with and without the application of a Gaussian filter.

\subsection{Validation of the NN potential}
\begin{figure*}[!h]
            \centering
            \includegraphics[width=0.9\textwidth]{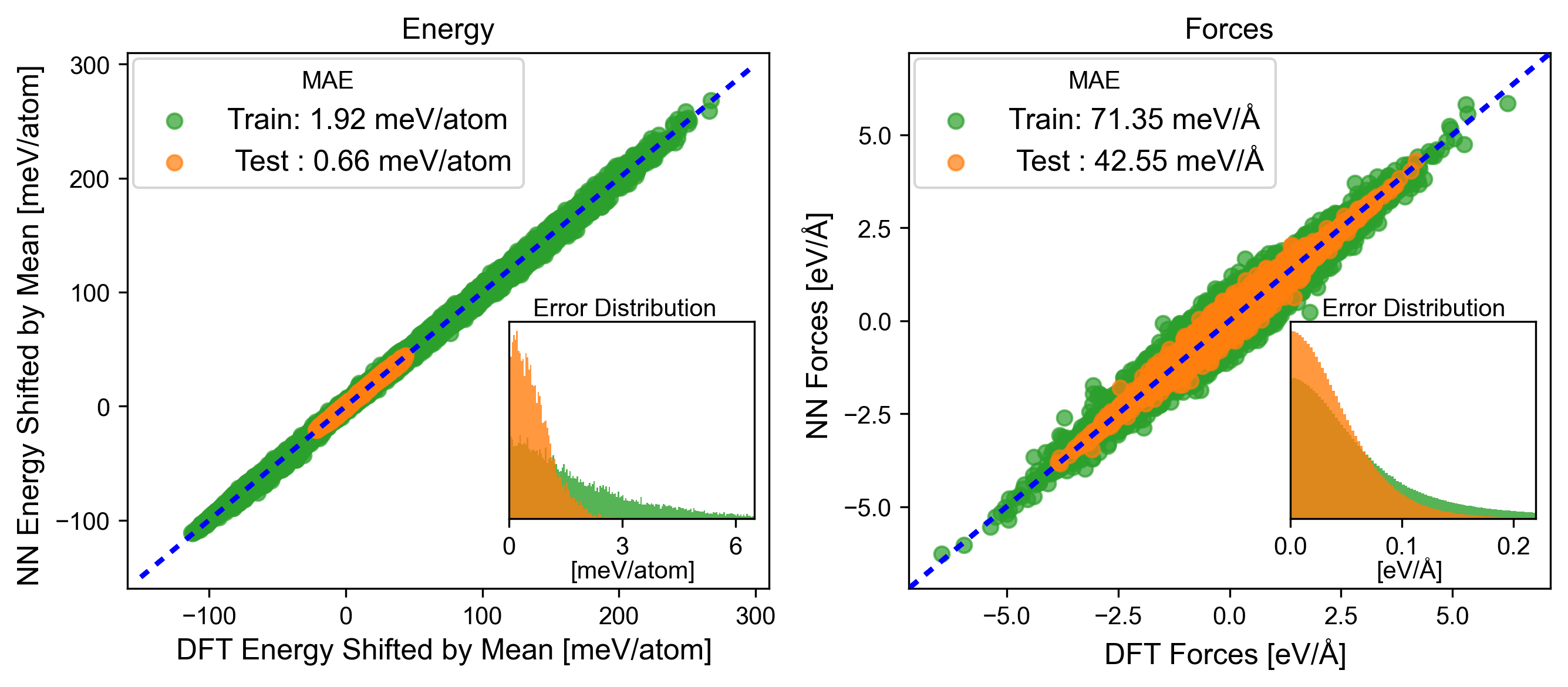}
            \caption{Comparison of the energies (left panel) and atomic forces (right panel) calculated on the training set (green) and test set (orange) using DFT and the final NN potential. Energies in the left panel are shifted by the mean value of the DFT atomic energies. Insets illustrate the probability distributions of the absolute difference in energies and forces between the DFT and final NN model}
            \label{supp_fig:NN_validation}
        \end{figure*}

        The mean absolute error (MAE) of energies in the training and test set are 1.92 and 0.66 meV/atom, respectively. The MAE of forces in the training and test set are 71.35 and 42.55 meV/{\AA}, respectively (see Fig.~\ref{supp_fig:NN_validation})
        The test set was composed of roughly 10,000 atomic configurations collected from both the unbiased simulations for polymer and molecular phases, as well as from the biased simulations in which the active process of the polymerization of S$_8$ sulfur and its reverse were observed. 
        These simulations were performed using our final NN potential model on systems made of 512 atoms. The temperature was set to 432 $\sim$ 500 K, and the simulation box was set to 24.8 {\AA} to be consistent with the experimental densities. 
        The comparison of the DFT and NN predicted energies and atomic forces over the training set and the test set are given in Fig.\ref{supp_fig:NN_validation}. 
        It is not strange that the MAE of the training set is larger than that of the test set because many active configurations in high temperature and pressure ranges were included in the training set, which thus displays a broader range of energies and forces distribution.

\clearpage
                    
\section{Supplementary Figures}
       \begin{figure*}[!h]
            \centering
            \includegraphics[width=0.75\textwidth]{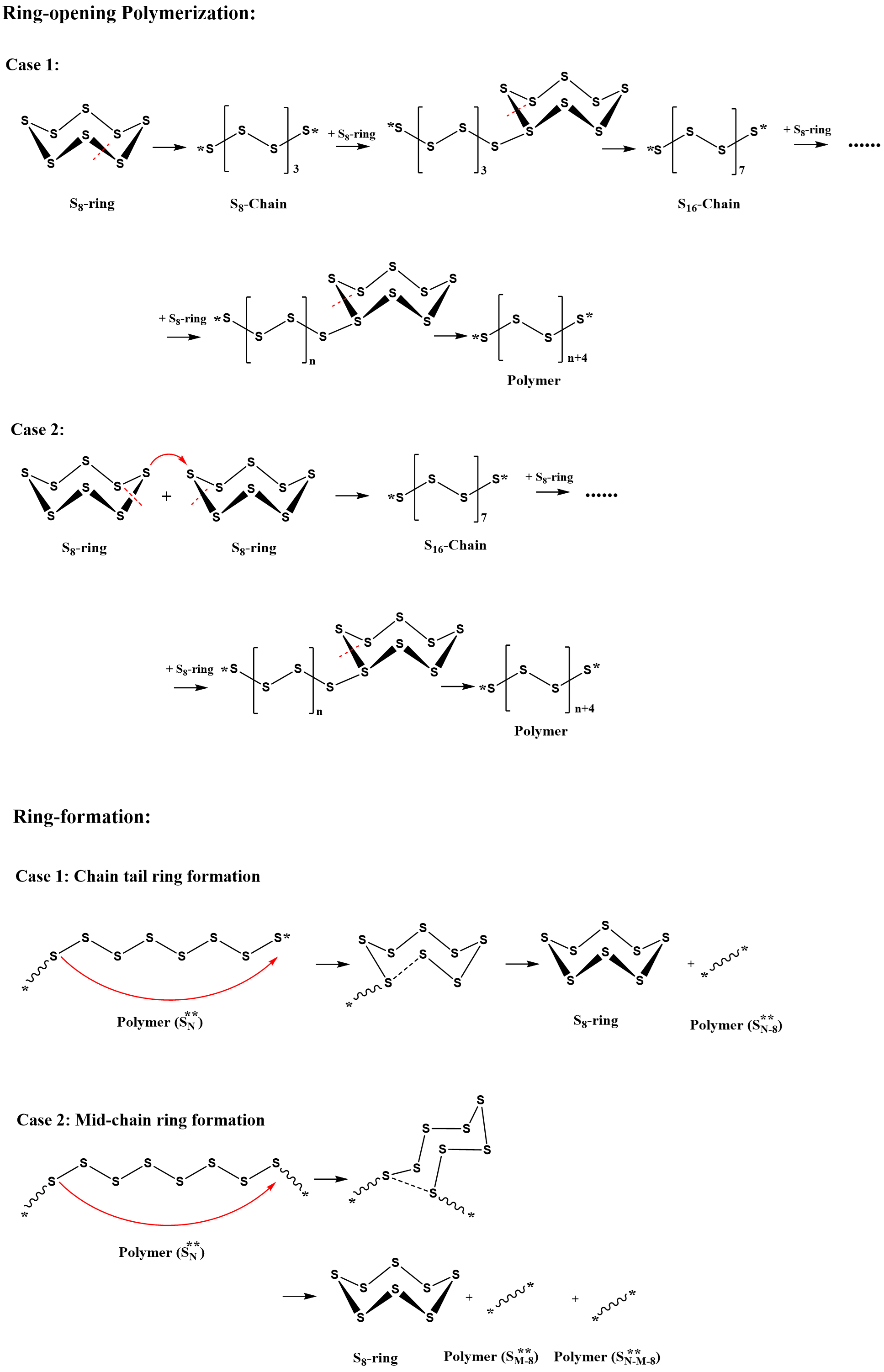}
            \caption{Schematic chemical representation of the polymerization and depolymerization mechanisms described in the main text.}
            \label{figure1}
        \end{figure*}
\clearpage
         \begin{figure*}[!h]
            \centering
            \includegraphics[width=0.9\textwidth]{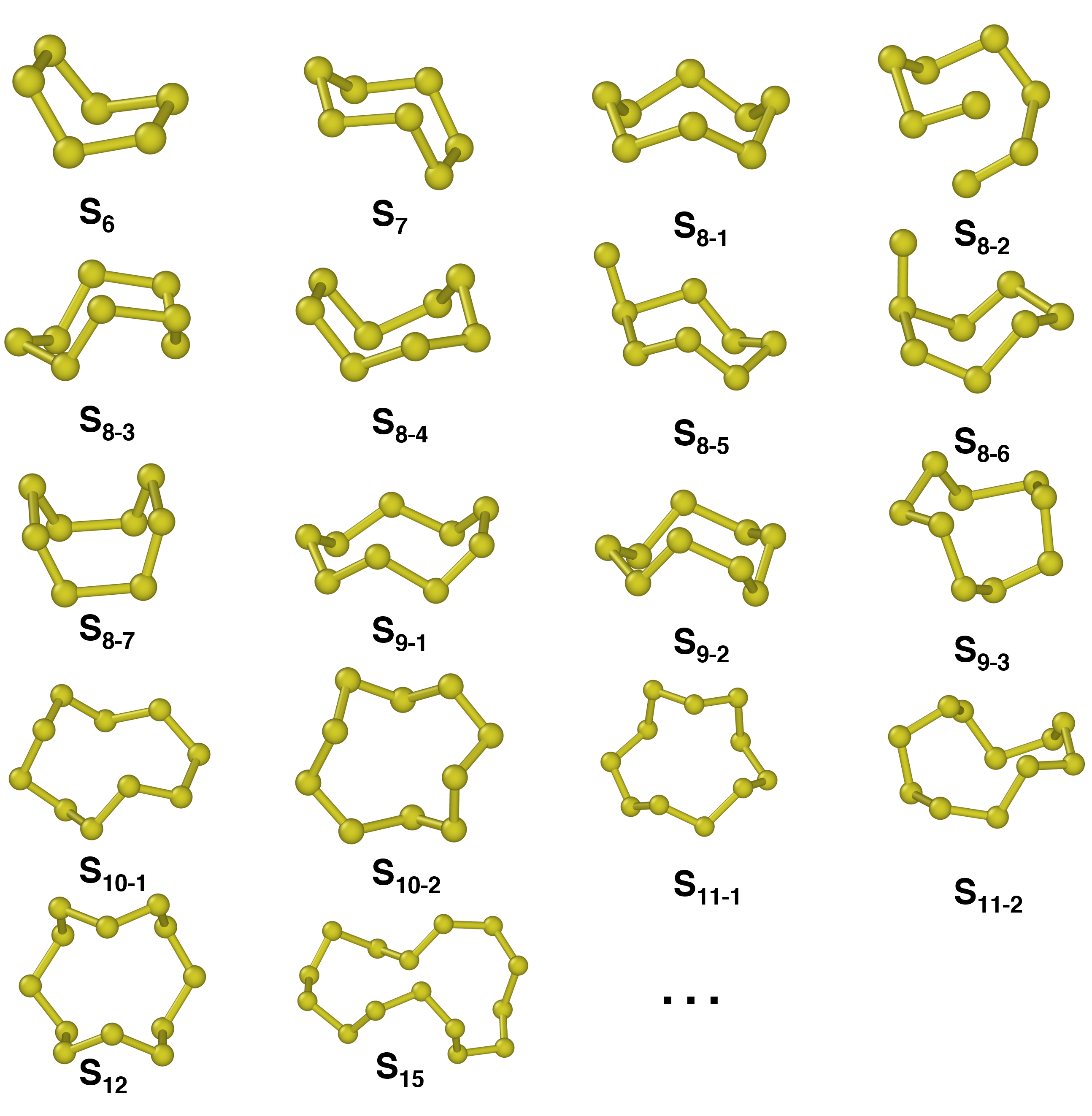}
            \caption{Snapshots of typical cyclic configurations (6$<$ n $<$ 15 ) observed in our simulations. Such configurations were already reported in the literature and validated with DFT calculations. See Ref.~\cite{steudel2003elemental} and the references hereby.}
            \label{supp_fig:Sn_rings}
        \end{figure*}  
\clearpage        
        \begin{figure*}[!h]
            \centering
            \includegraphics[width=0.9\textwidth]{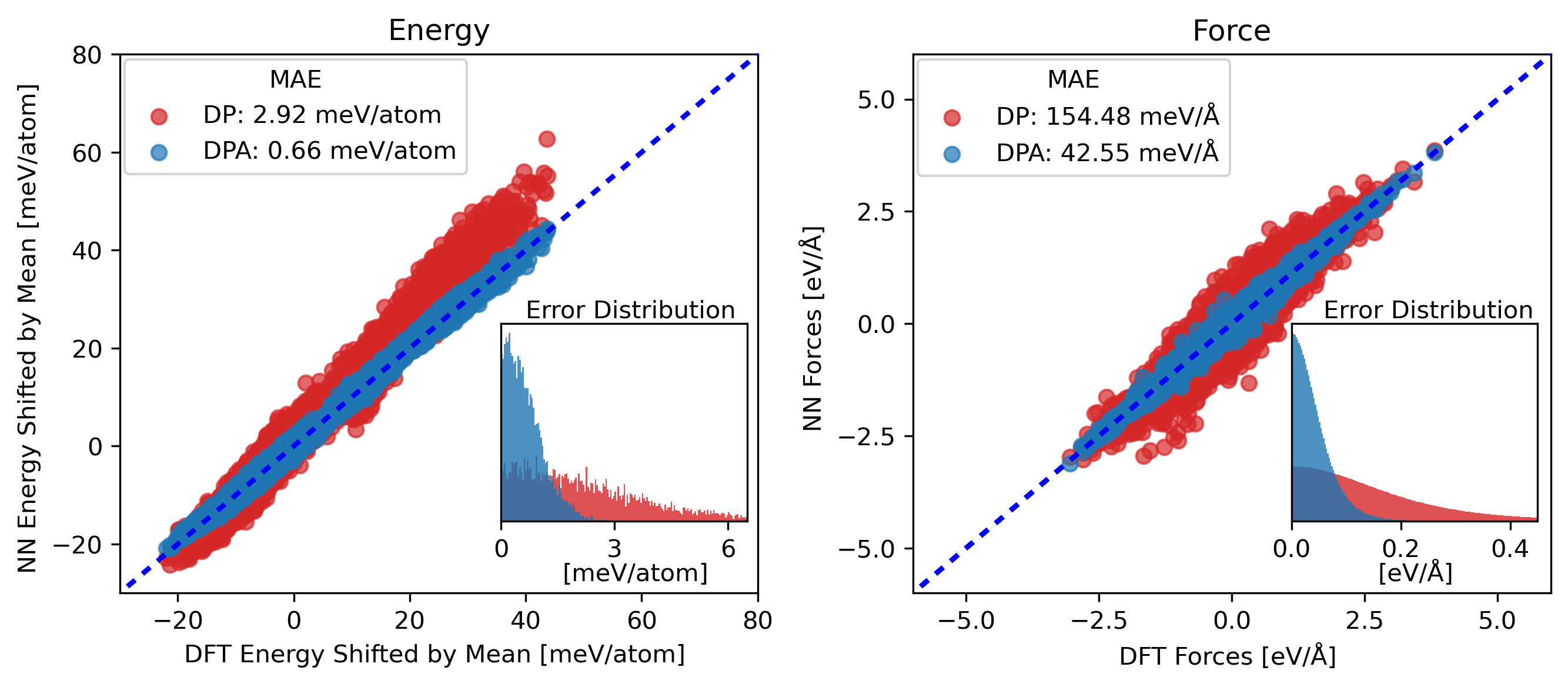}
            \caption{Comparison of energies (left panel) and atomic forces (right panel) calculated on the test set using DFT and NN potentials trained with the Attention-based Deep Potential scheme (DPA, blue) and the standard Deep Potential-Smooth Edition scheme(DP, red). Energies in the left panel are shifted by the mean value of the DFT atomic energies. Insets illustrate the probability distributions of the absolute difference in energies and forces between the DFT and NN models. For the standard DP potential, we trained on the same training set as that of our final NN potential (DPA), as well as the same hyperparameters except for increasing the training steps from 3 $\times$ 10$^6$ to 5 $\times$ 10$^6$. The test is the same as that used to validate our final NN potential.}
            \label{supp_fig:NN_comparison}
        \end{figure*}
\clearpage

\bibliography{references}